Manuscript accepted by JSWSC

# Unseasonal super ionospheric plasma bubble and scintillations seeded by the 2022 Tonga Volcano Eruption related perturbations

Wenjie Sun[1,2,3,4], Ajith K. K.[5], Guozhu Li[1,2,3,4*], Yu Li[6], Xiukuan Zhao[2,3,7], Lianhuan Hu[1,2,3], Sipeng Yang[1,2,3], Haiyong Xie[1,2,3], Yi Li[1,2,3], Baiqi Ning[1,2,3], and Libo Liu[2,3,4,7]

[1] Beijing National Observatory of Space Environment, Institute of Geology and Geophysics, Chinese Academy of Sciences, Beijing, China.
[2] Key Laboratory of Earth and Planetary Physics, Institute of Geology and Geophysics, Chinese Academy of Sciences, Beijing, China.
[3] Innovation Academy for Earth Science, Chinese Academy of Sciences, Beijing, China.
[4] College of Earth and Planetary Sciences, University of Chinese Academy of Sciences, Beijing, China.
[5] National Atmospheric Research Laboratory, Gadanki, India.
[6] China Earthquake Networks Center, China Earthquake Administration, Beijing, China.
[7] Mohe Observatory of Geophysics, Institute of Geology and Geophysics, Chinese Academy of Sciences, Beijing, China

***Corresponding author: Guozhu Li (gzlee@mail.iggcas.ac.cn)**

## Abstract

The Hunga-Tonga volcano eruption at 04:14:45 UT on 15 January 2022 produced various waves propagating globally, disturbing the background atmosphere and ionosphere. Coinciding with the arrival of perturbation waves, several equatorial plasma bubbles (EPBs) were consecutively generated at post-sunset hours over the East/Southeast Asian region, with the largest extension to middle latitudes. These EPBs caused intense L-band amplitude scintillations at middle-to-low latitudes, with signal fading depths up to ~16 dB. Considering the very rare occurrence of EPBs during this season in East/Southeast Asian sector and the significantly modulated background ionosphere, we believe that the perturbation waves launched by the volcano eruption triggered the generation of the unseasonal super EPBs. The ionospheric perturbations linked with the 2022 Tonga volcano eruption propagated coincidently through the East/Southeast Asia longitude sector near sunset, modulated the equatorial F region bottomside plasma density and acted as the seeding source for the generation of the unseasonal super bubbles. Our results implicate that volcano eruption could indirectly affect the satellite communication links in the region more than ten thousand kilometers away.

# 1. Introduction

Volcano eruption is one of the deadly disasters which can affect both the solid earth and the upper atmosphere, associated with numerous hazards including such as earthquakes, tsunamis, lightning and air pollution (e.g., Schmidt et al., 2014). The broad scientific and socioeconomic effects linked with large volcanic explosion are of great research value in many areas. In the field of space weather, the sudden injection of energy and momentum into the atmosphere due to volcanic eruptions could produce various waves, propagating long distances and disturbing the background atmosphere and ionosphere (e.g., Cheng and Huang, 1992). The most recent Hunga-Tonga volcano eruptions on 15 January 2022 is considered as the most significant explosion in the last 30 years, releasing enormous energy and inducing globally propagating waves disturbing the ionosphere to a large extent which was reported in several recent papers (e.g., Lin et al., 2022; Zhang et al., 2022).

Equatorial plasma bubble (EPB) is thought of as a Rayleigh-Taylor (R-T) instability process seeded by the perturbation waves at the ionospheric F region bottomside. It is believed that the growth rate of R-T instability could be enhanced by driving forces such as the pre-reversal enhancement of the eastward electric field (PRE), thus control the generation of EPBs (e.g., Kelley, 2009; Abdu, 2019). Once developed, EPBs could rise to the topside ionosphere via the polarization electric field inside, elongating along the magnetic field lines and mapping to the latitudes away from the equator (Otsuka et al., 2002; Keskinen et al., 2003). Strong EPBs may cause signal fading, scintillation, and even severe outages in the satellite-based communication and navigation systems in a large region (e.g., Seo et al. 2009; Alfonsi et al., 2013). So it is of vital importance to figure out the variation of EPB occurrences to mitigate possible risks therefrom. However, the day-to-day variability in the occurrence of EPBs is not yet fully understood due to the sophisticated background atmospheric and ionospheric conditions, seeding sources and driving forces.

In the East/Southeast Asia, EPB is mainly an equinoctial phenomenon at low latitudes due to the specific magnetic field configuration, and was seldom observed previously during the winter months of the northern hemisphere (e.g., Buhari et al., 2017; Li et al., 2021; Zhao et al., 2021). But under certain conditions when the seeding source and/or the driving forces were satisfied, unseasonal EPBs may also occur. For example, during the main phase of the severe geomagnetic storm on 12 February 2000, super EPBs were observed extending to middle latitudes that was attributed to the effect of storm time prompt penetration electric field (e.g., Ma and Maruyama, 2006). In this paper, we report an unusual case of unseasonal super EPBs extending to middle latitudes on 15 January 2022. The EPBs caused strong ionospheric scintillations and satellite signal fading at middle and low latitudes. The onsets of EPBs coincided well with the arrival of the waves induced by the Tonga volcano eruption. The waves produced significant ionospheric perturbations which could act as a strong seeding source for the generation of unseasonal EPBs.

# 2. Data and Methods

A combination of Global Navigation Satellite System (GNSS) measurements from the Crustal Movement Observation Network of China (CMONOC) (Aa et al., 2015), the Ionospheric Observational Network for Irregularity and Scintillation in East/Southeast Asia (IONISE) (Li et al., 2019; Sun et al., 2020), the International GNSS Service (IGS) Working Group on ionosphere (Beutler et al., 1999), and the Chinese Meridian Project (Wang, 2010) were employed in this study. The vertical TEC were calculated using the same method as in previous studies (e.g., Xiong et al., 2016). The time resolution of TEC data used in the present analysis is 30 s. The 60-min running average from each vertical TEC time series was subtracted to obtain the background ionospheric perturbations. The Rate of TEC index (ROTI; Pi et al., 1997) were calculated every 5 min to characterize the kilometer-scale irregularities associated with EPBs. The latitudes/longitudes of ionospheric pierce points (IPPs) were calculated by assuming the F-region height at 300 km. The 50-Hz raw data were recorded at several GNSS sites of the IONISE network. The signal fading depth and the 1-min resolution amplitude scintillation index $S_4$ were calculated from the raw data. The 5-min running average of the power was subtracted for each tracked satellite and the absolute values were taken as the signal fading depth. For all the observations, only the data from the satellites with elevation angles lager than 30°were used. In addition, the in-situ electron density ($N_e$) profiles obtained from the Swarm satellites (Friis-Christensen et al., 2008; Xiong et al., 2019) were employed to investigate the occurrence of EPBs.

The ionosonde ionograms obtained at Beijing (116.2°E, 40.3°N, dip Lat, 39.4°), Wuhan (114.5°E, 31.0°N, dip Lat, 28.0°) and Ledong (109.0°E, 18.4°N, dip Lat, 12.9°) with temporal resolutions of 15 min, 15 min and 1 min respectively were manually scaled to get the background ionospheric conditions at different latitudes. The ΔH measurements derived from the magnetometers at the stations DLT (108.5°E, 11.9°N, dip Lat, 5.1°) and Ledong, which are near and away from the magnetic dip equator respectively, were employed to indicate the equatorial electric field (Anderson et al., 2002). To make a comparison between the normal day and the case day, the monthly mean peak height of $F_2$ layer ($h_mF_2$) and ΔH in January 2022 (14-16 January were excluded) were calculated. Also, the brightness temperature measurements in the 4.3μm radiance which were obtained from the Atmospheric Infrared Sounder (AIRS) aboard NASA's Aqua satellite were employed to indicate the stratospheric gravity wave activity (e.g., Hoffmann et al., 2013).

# 3. Results

On the Tonga volcano eruption day 15 January 2022, strong L-band amplitude scintillations and signal fading were observed in a broad region over the East/Southeast Asian sector during post-sunset hours. Figure 1 shows the $S_4$ index and signal fading depth projected on geographic maps (top panels) with data during 10:30-18:00 UT, and at some selected sites (bottom panels) during 07:00-21:00 UT. Strong scintillation and signal fading were observed at low and middle latitudes (up to ~35°N), with $S_4$ index (signal fading depth) up to 0.8 (16 dB) around 20-30 °N. The observations in Figure 1 indicate the possible occurrence of EPBs over a broad region. It is well-known that EPBs are embedded with irregularities of various scales, and such strong L-band scintillations are mainly produced by hundred-meter-scale irregularities (Basu et al., 1988). Under strong scintillations caused by super bubbles, the loss-of-lock of GNSS signals and positioning errors could occur even at middle latitudes (e.g., Demyanov and Astafyeva, 2012). For the present case, severe scintillation and signal fading were caused even at the latitudes higher than 30 °N. The strongest scintillation and signal fading depth were observed near the equatorial ionization anomaly region, where the background plasma density is usually higher than that at other latitudes. The larger density gradient in this latitude region would be favorable for producing stronger scintillations under the presence of EPBs (e.g., Abadi et al., 2014).

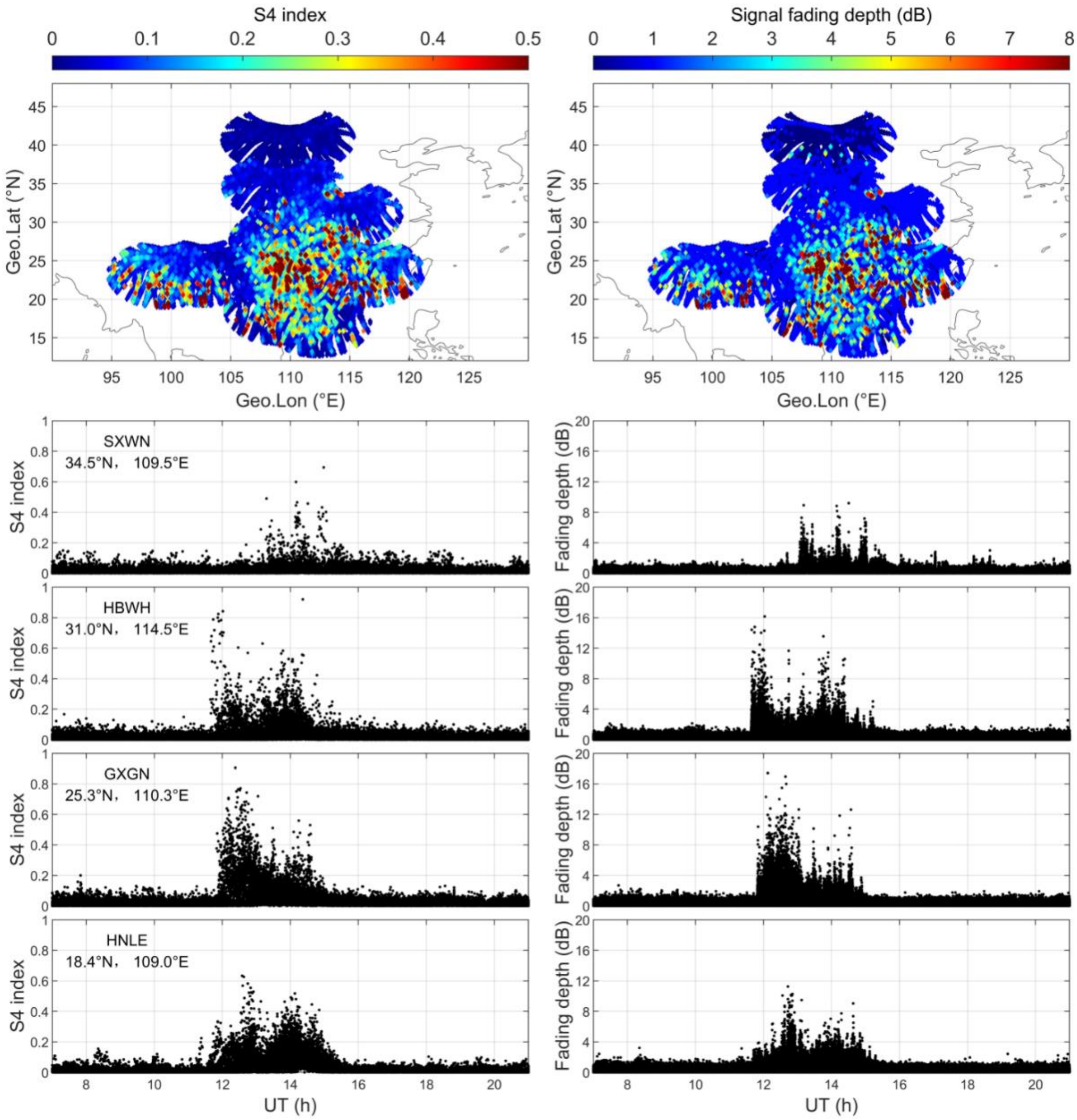

***Figure 1.*** *The $S_4$ index (left panels) and signal fading depth (right panels) projected on*

*geographic maps (top panels, using data during 10:30-18:00) and at selected sites (bottom panels) on 15 January 2022.*

In order to investigate the morphology and evolution of the EPBs causing the strong scintillation, Figure 2 shows a sequence of ROTI maps during 10:15-14:30 UT. Four main features can be seen from Figure 2: (1) No EPB was detected before sunset (i.e., at 10:15 UT). (2) As the sunset terminator passed, one EPB (named as bubble “A”) was generated in the longitudes 120-125 °E around 10:35 UT and then drifted westward. At 12:00 UT, another EPB (named as bubble “B”) at the east of bubble “A” can be roughly distinguished, which was likely generated earlier than bubble “A” at the more eastern longitudes and drifted westward into the map. As the two bubbles drifted zonally, they also extended to higher latitudes gradually, indicating the growth of plasma bubbles from lower to higher altitudes at the magnetic equator. At 12:30 UT, the northern edge of the two bubbles reached ~28 °N (dip Lat, 24.5 °) and ~33 °N (dip Lat, 30.8 °) respectively. (3) A new EPB (named as bubble “C”) was freshly generated over ~100 °E at 12:20 UT and grew up gradually. After 12:55 UT, an east-tilted branch with large ROTI values was initially observed. The branch, which grew quickly and extended to the latitude ~48 °N at 13:15 UT, was unlikely a part of the EPBs (without coincident scintillations). (4) The bubbles started decaying after 13:15 UT, nearly stopped drifting and persisted until 18:00 UT.

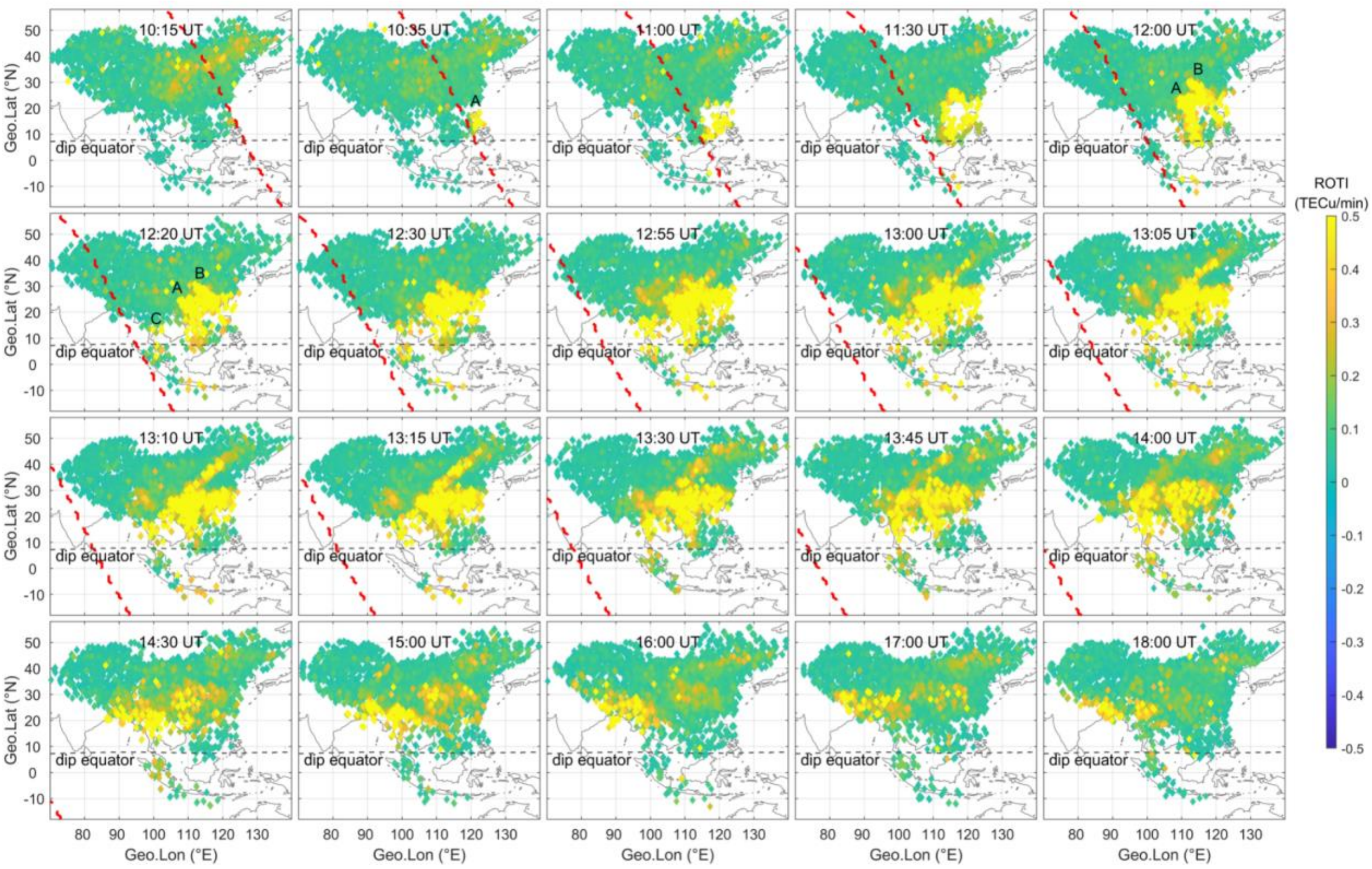

***Figure 2.*** *A sequence of ROTI maps from 10:15-18:00 UT on 15 January 2022 observed by GNSS networks showing the evolution of EPBs. The black and red dashed curves in each panel denote the magnetic dip equator and the sunset terminator at E-region altitude (100 km) respectively.*

During the occurrence of the EPBs, the Swarm satellites B and C happened to fly over the East/Southeast Asian sector. Figure 3 shows two $N_e$ profiles recorded by Swarm C and Swarm B satellites along 99°E and 108°E at the altitudes of 450 km and 510 km during 16:10-16:40 UT and 16:00-16:30 UT, respectively. From Figure 3, significant density fluctuation and depletion were observed over the equatorial and low-latitude regions (10°S-25°N) along 99°E, while only weak density fluctuation and depletion were observed over limited latitudes (25-35°N and 10-18°S) along 108°E. It should be noted that during the periods of the two Swarm satellites paths the EPBs were at the decaying stage, when the intensity weakened near 108°E but still persisted near 99°E (Figure 2). Nevertheless, the Swarm $N_e$ profiles correspond well with the ROTI maps, confirming the occurrences of EPBs.

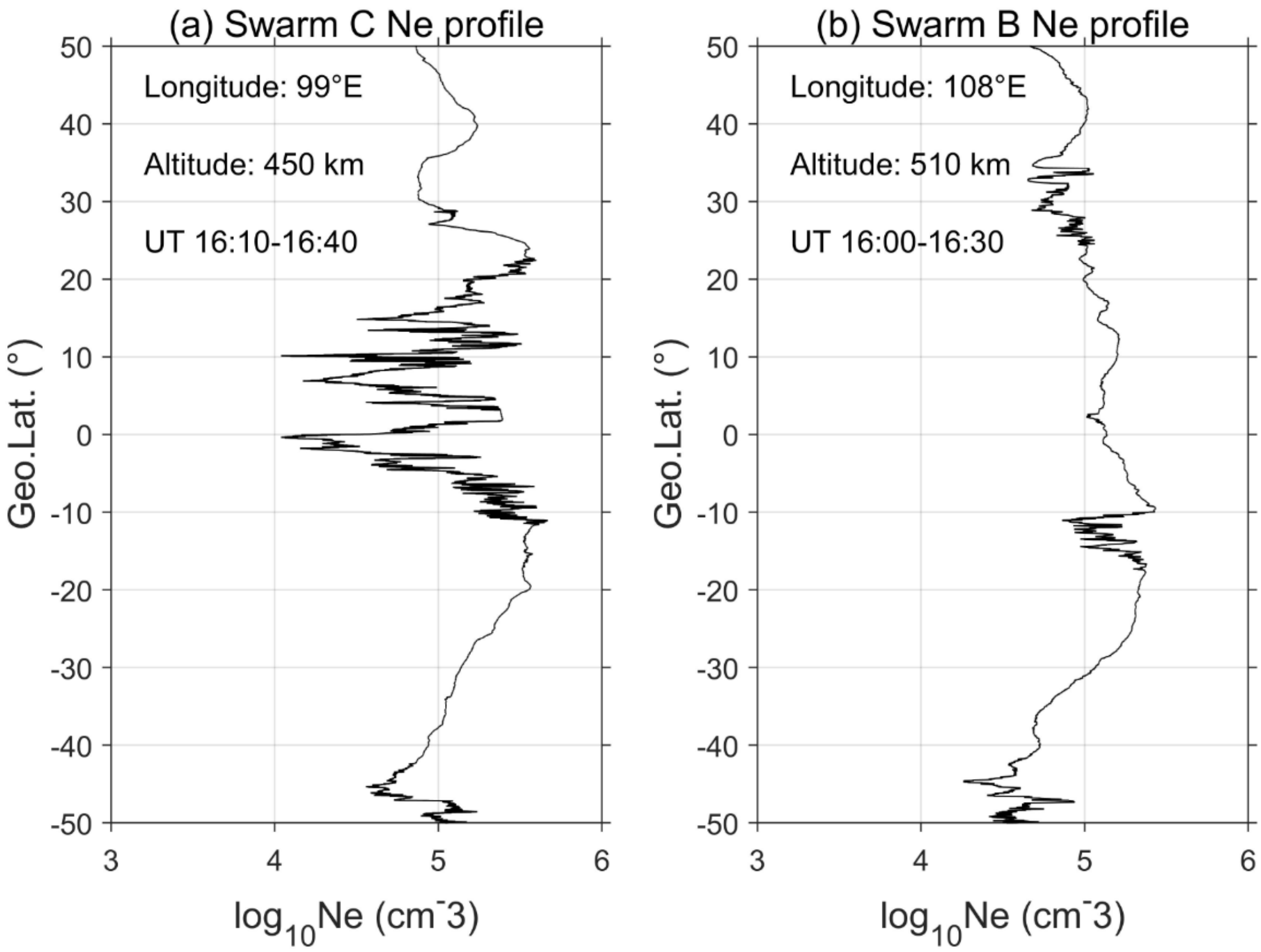

***Figure 3.*** *The in-situ plasma density profiles obtained from (a) Swarm C and (b) Swarm B satellites along 99°E and 108°E at the altitudes of 450 km and 510 km during 16:10-16:40 UT and 16:00-16:30 UT, respectively.*

The generation processes of bubbles "A" and "C" were completely recorded by the ROTI maps. The generation of bubble "B" was likely the earliest, which failed to be detected due to the data gaps over the oceans. The bubbles "A" and "B" were slightly west-tilted. Compared with previous studies, the bubbles in this case are of particular interest in three aspects. (1) Over the Asian sector, EPBs prefer to occur in equinox of solar maximum (e.g., Shi et al., 2011; Buhari et al., 2017). However, this case occurred in winter months of the northern hemisphere and the solar activity was at the early rising stage from solar minimum. (2) The drift of EPBs is usually eastward (e.g., Fejer et al., 2005). However in the present case, it was westward. (3) EPB is mainly confined at low latitudes within ±20 ° latitude under solar minimum years in this longitude sector (Li et al., 2021). However, the EPBs in this case extended to middle latitudes of more than 30 °N that can be termed as super bubbles.

## 4. Discussion

EPBs are usually thought to be generated due to the R-T instability (Kelley, 2009). There are two major factors controlling the generation of EPBs, the seeding perturbation source for R-T instability at the F region bottomside, and the driving force for the growth of the R-T instability (Abdu, 2019). In the East/Southeast Asian sector, EPBs were usually observed in equinoctial months when the PRE, which is considered as a major driving force, are stronger during this period as the sunset terminator is aligned with the magnetic meridian (e.g., Buhari et al., 2017; Zhao et al., 2021). However, the present case was observed during the northern hemispheric winter, when the condition was usually unfavorable for the generation of EPBs.

To investigate possible mechanisms responsible for the present super bubbles, Figure 4 shows the observations of background parameters, the interplanetary magnetic field (IMF) $B_z$, the interplanetary electric field (IEF) $E_y$, Dst and ΔH ($H_{DLT}$-$H_{Ledong}$), and the $h_mF_2$ obtained from three ionosondes from middle to low latitudes during 14-15 January 2022. The grey dashed curves represent the monthly mean values. The shaded areas mark the approximate period of EPBs. It can be seen that a moderate geomagnetic storm occurred on 14 January. The super bubbles occurred during the geomagnetic storm later recovery phase. As indicated by the variation of ΔH, the equatorial electric field during the daytime of 15 January was predominantly westward (negative). Notably, the ΔH began to increase since ~09:00 UT and turned positive at ~09:40 UT, ~1 hour before the first EPB was detected. Corresponding with the increase of ΔH, the F layer at different latitudes was elevated gradually (as characterized by $h_mF_2$ increase). The rising height around 10:30-11:30 UT was 80-130 km. The elevated F layer would provide favorable conditions for the growth of R-T instability and thus the generation of EPBs (e.g., Kelley, 2009; Abdu, 2019).

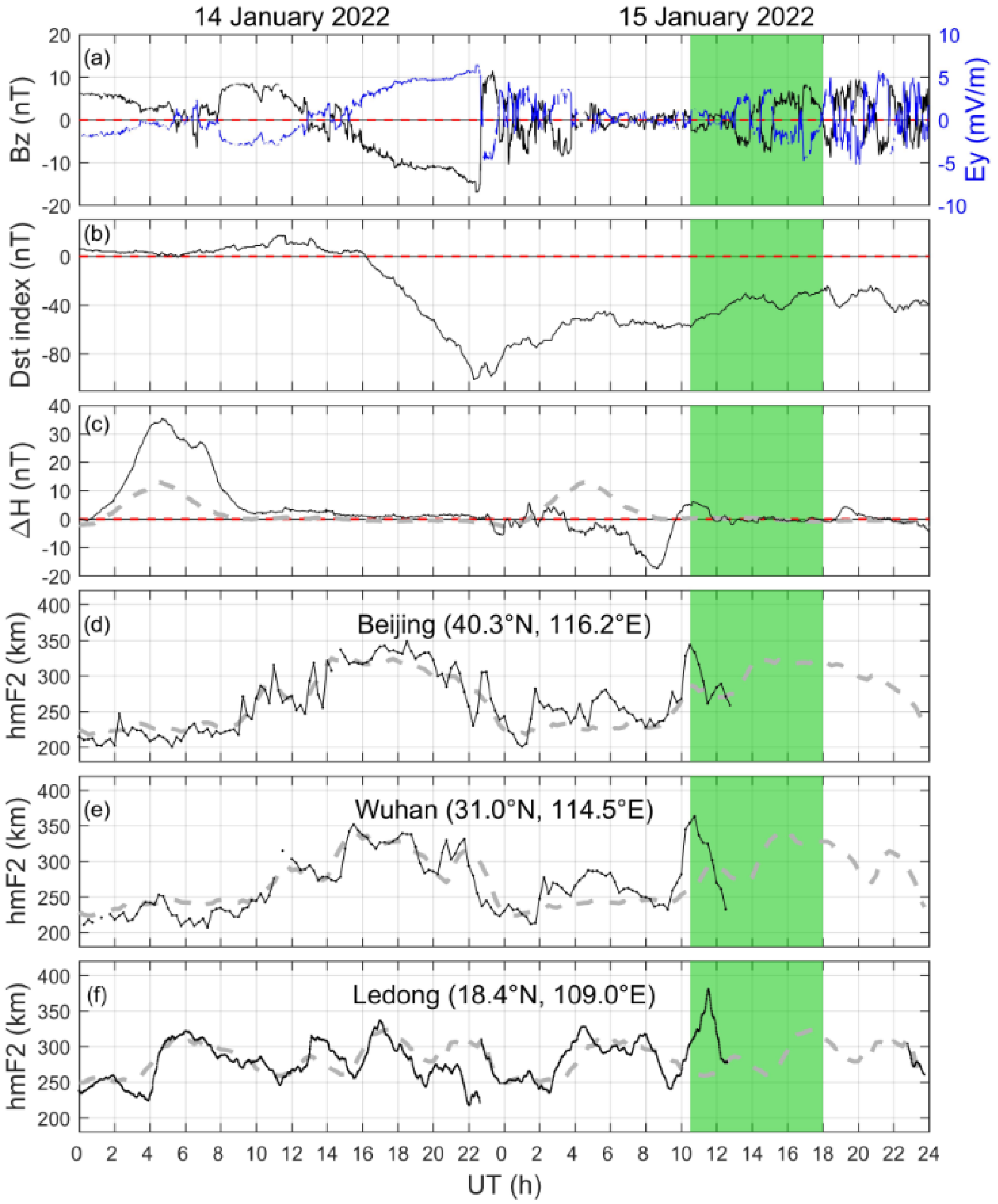

***Figure 4.*** *The (a) IMF $B_z$ (black curve) and IEF $E_y$ (blue curve), (b) Dst, (c) $\Delta H$, and $h_mF_2$ observed at (d) Beijing, (e) Wuhan and (f) Ledong during 14-15 January 2022. The green shaded areas mark the approximate period of the EPBs. The grey dashed curves represent the monthly mean values. The red horizontal dashed lines in (a-c) mark the value of zero.*

Regarding the factors responsible for the elevation of F layer around sunset, one possibility is storm time electric field. The super bubbles extending to middle latitudes were previously observed during strong geomagnetic storms when the F layer could be significantly elevated by storm time prompt penetration electric field (e.g., Ma and Maruyama 2006; Katamzi et al., 2017). For the present case, the EPBs were generated during the later recovery phase of the geomagnetic storm when the equatorial electro-dynamics are usually controlled by westward dynamo electric fields. The IMF $B_z$ and IEF $E_y$ remained relatively quiet during the period when the EEJ turned from westward to eastward (09:00-12:00 UT, Figure 4). Notably, previous studies reported considerable penetration electric fields even without severe IMF $B_z$ fluctuations (e.g., D'Angelo et al., 2021; Piersanti et al., 2020). More recently, Harding et al. (2022) proposed that if the penetration electric fields were the main cause of the EEJ variation, strong correlations between the IEF $E_y$ and ΔH would be expected. A correlation coefficient as high as 0.6 was observed in the Peru and Brazil sectors during the post-volcanic-eruption hours. However, Harding et al. (2022) also pointed out that it was the 1-hour-scale EEJ fluctuations that were likely caused in part by the penetration electric field, but the larger and longer perturbations were not. In the East/Southeast Asian sector, the ΔH variation during 09:00-12:00 UT was of a much larger scale than 1 hour (as shown in Figure 4). This was not obviously correlated with IEF $E_y$ fluctuations. The storm time electric field was unlikely the main reason for the elevation of F layer in the present case.

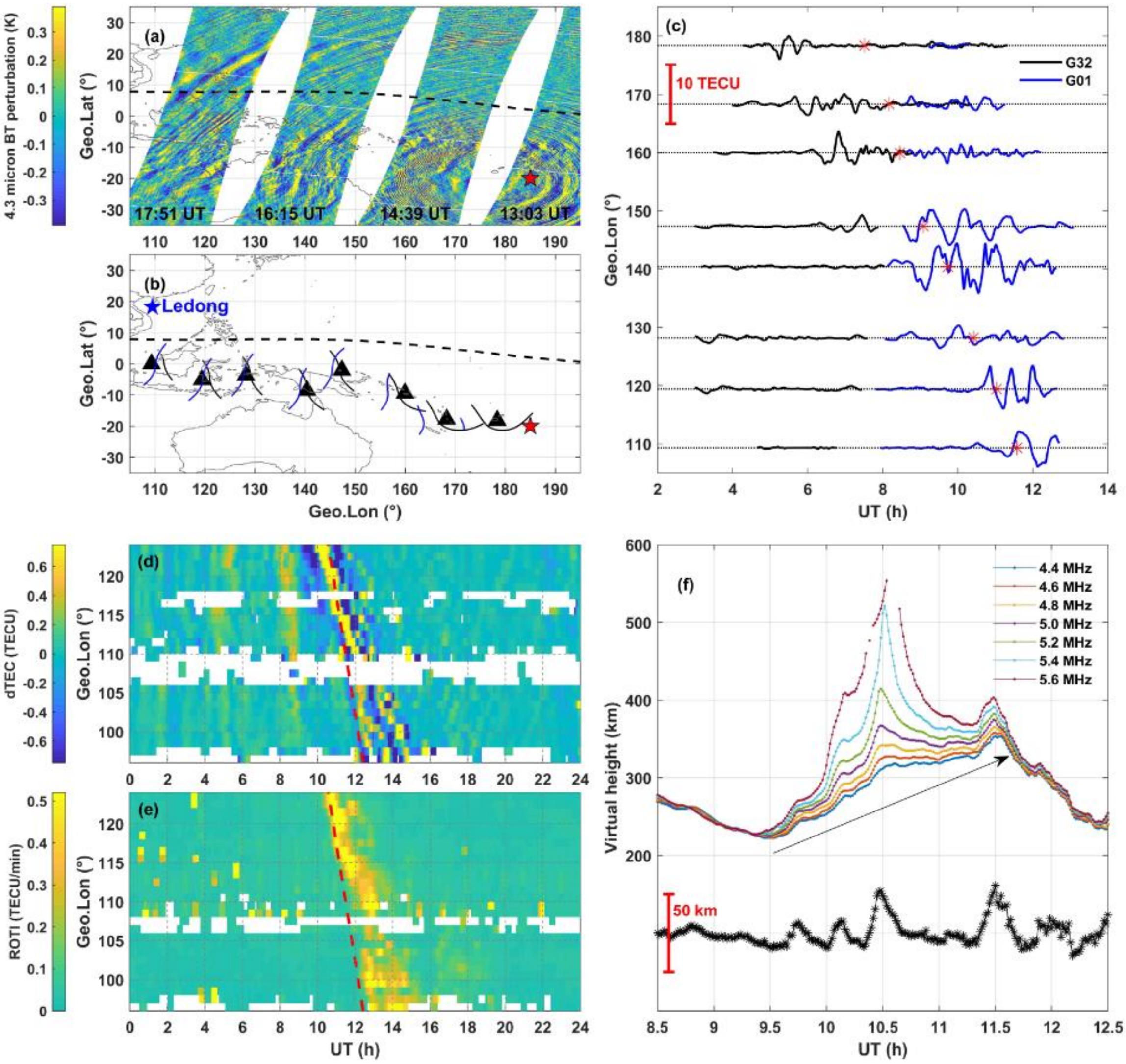

***Figure 5.*** *(a) AIRS observation of concentric stratospheric gravity waves in 4.3 μm radiance. (b) The geographic locations of GNSS sites (triangles) and IPP trajectories (black and blue solid curves) connecting Tonga to the magnetic dip equator conjugate to the East/Southeast Asia. (c) The dTEC obtained from the GPS satellites G01 and G32 for the GNSS sites in (b).*

*(d) The dTEC and (e) ROTI at the Asian longitudes along the magnetic dip equator averaged using all the data at 0-15 °N. (f) The plasma virtual heights at several selected frequencies derived from the Ledong ionosonde and the 60-min filtered virtual height at the plasma frequency 5.0 MHz (black asterisk curve). The red stars and black dashed curves in (a-b) mark the location of the Tonga volcano and the magnetic dip equator respectively. The blue star in (b) marks the location of Ledong. The red asterisks in (c) indicate the E-region sunset at the location of each site. The red dashed curves in (d-e) denote the E-region sunset terminator. The black arrow in (f) indicates the elevation of the F layer bottomside.*

Another possibility is the travelling ionospheric disturbance (TID) associated polarization electric field. It was suggested that the TID/gravity waves (GWs) rising from the tropospheric origins could induce zonal polarization electric fields and favor EPB development regardless the presence of PRE (see Figure 1 of Abdu, 2019). In this regard, the perturbation waves linked with the Tonga volcano eruption on 15 January 2022, which are manifested as TIDs at the ionosphere altitude, have been reported in the very recent papers (e.g., Lin et al., 2022; Zhang et al., 2022). These TIDs were suggested to travel at the sound speed ~340 m/s and arrived at the Asian sector ~6 hours after the eruption (~11:00 UT). Figure 5a shows the AIRS observation of concentric stratospheric GWs. Note that the observations were a composite of different images taken at different times during ~13:00-18:00 UT. The average time stamp of each AIRS orbit is marked at the bottom of each snapshot. The plot shows clearly the GWs with epicenter at the Tonga volcano, which can reach the Asian longitudes even after 13 hours of the volcanic eruption. Figures 5b-c show the detrended TEC (dTEC, calculated by subtracting the 60-min running average) observed from the satellites G01 and G32 at 8 GNSS sites along the path approximately between Tonga and the magnetic dip equator at the Asian longitudes. Figures 5d-e present the dTEC and ROTI at the Asian longitudes along the magnetic dip equator, averaged using all the data at 0-15 °N (which is within the area ±7.5 ° astride the magnetic dip equator). It can be seen from Figure 5c that wave signatures were detected sequentially along the passage from Tonga to Asia after the volcano eruption at 04:14:45 UT. When the perturbations arrived at the equatorial ionosphere in the Asian longitudes around sunset 10:00 UT, the EPBs were then observed afterwards. Figure 5f shows the plasma virtual heights at several selected frequencies derived from the Ledong ionosonde. The virtual height at the plasma frequency 5.0 MHz after subtracting the 60-min running average was also plotted as the black asterisk curve. It can be seen that the F layer was elevated during 09:30-11:30 UT (indicated by the black arrow), with increasing and periodically oscillating virtual heights. The observation indicated that besides the background electric field, the F layer might be further elevated by the TID-induced polarization electric field.

Regarding the seeding mechanism, it was suggested that the TID-associated F layer height oscillations in the form of large-scale wave structures getting amplified towards sunset could seed the development of the R-T instability, which always appear to be the precursor to post-sunset EPBs (e.g., Abdu et al., 2019; Takahashi et al., 2018). From Figure 5, the arrival of the eruption-produced perturbation waves at the Asian sector was about 20-30 minutes earlier than the onset of the EPBs detected in Figure 2. The TIDs arrived 120 °E and 100 °E at

~10:00 UT and ~12:00 UT respectively and thus could act as the seeding perturbation source. The growth rate of R-T instability could then be further enhanced by the eastward polarization electric field within the perturbation structure through elevating the F layer to higher altitudes, thus generating the super bubbles. It is relevant to mention that the storm-induced TIDs originated from the polar region that propagated southward could also be a potential seeding source for EPBs. In this regard, Figure 6 shows the TEC and dTEC (calculated by subtracting the 60-min running average) results obtained from the Beidou geostationary satellites along the 110°E receiver chain of IONISE (Sun et al., 2020). Each TEC and dTEC curve was offset by the corresponding latitude of the IPP. From Figure 6, there were no considerable TEC fluctuations propagating southward from higher latitudes to the equatorial region before 11:00 UT. During 11:00-15:00 UT, the fluctuations started nearly simultaneously at different latitudes (or a little earlier at lower latitudes than higher latitudes). The results indicate that no TIDs were observed propagating southward.

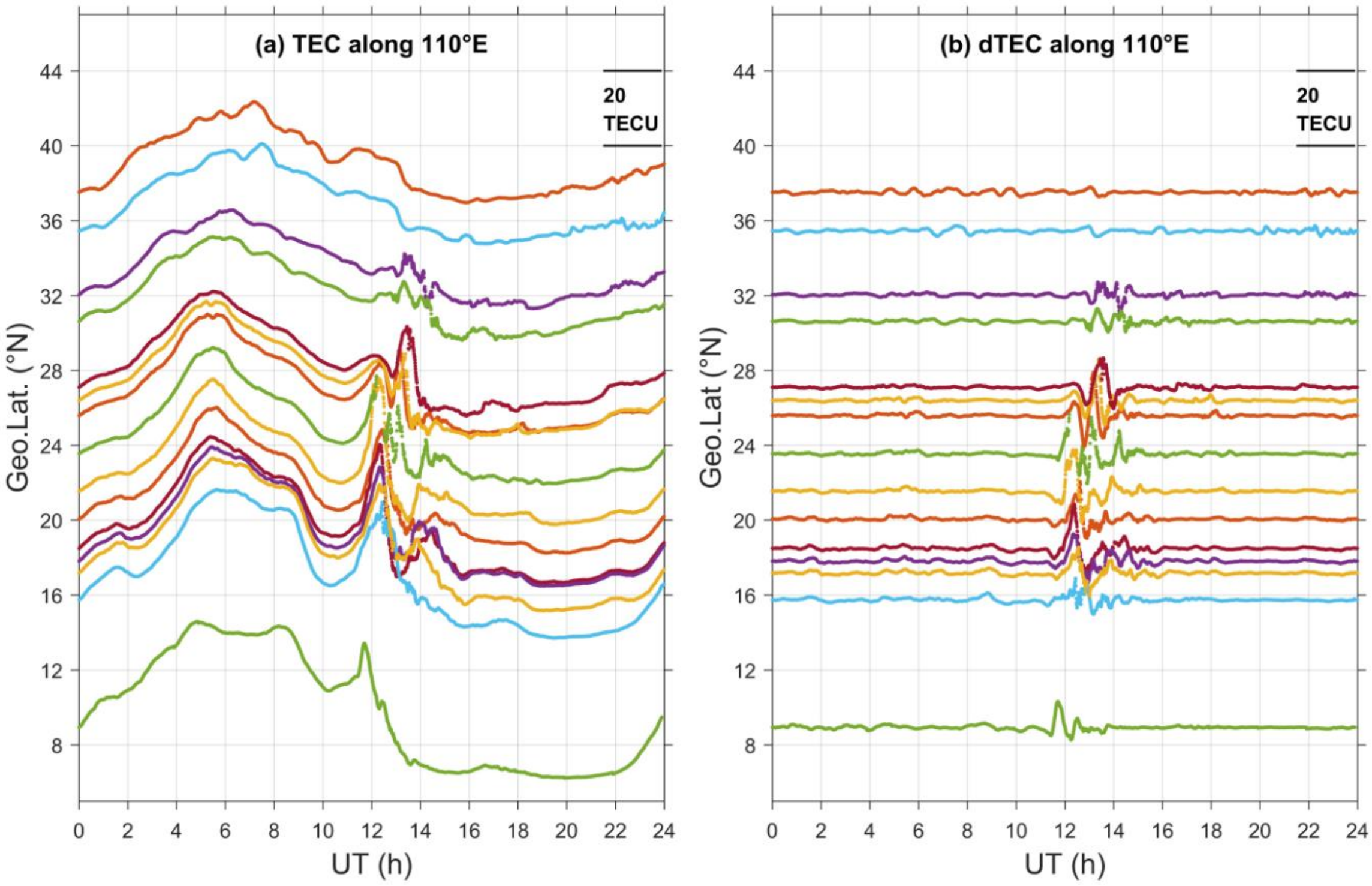

***Figure 6.*** *(a) TEC and (b) dTEC obtained from Beidou geostationary satellites along 110°E on 15 January 2022. Each TEC and dTEC curve was offset by the corresponding latitude of the ionospheric pierce point.*

On the other hand, one may argue that the irregularities characterized as ROTI enhancements observed in middle latitudes up to ~33°N (Figure 2) could possibly be locally triggered by the arriving TIDs via the Perkins instability (e.g., Alfonsi et al., 2013; Balan et al., 2018) rather than the extension of EPBs. Whereas the irregularities due to the Perkins instability are able to induce ROTI enhancements, they are unlikely to induce severe amplitude scintillations (Li et al., 2021). For the present case, severe amplitude scintillations were observed extending to middle latitudes (Figure 1). On the other hand, if these irregularities were associated with the arriving TIDs linked with the Tonga volcano eruption, they should elongate in the similar orientation with the wave front, which was supposed to be in the northeast-southwest direction in the East/Southeast Asia (Lin et al., 2022). However, the ROTI structures corresponding the onset of EPBs (especially structures “A” and “B”) were west-tilted that is different from the direction of the TID wave front. Further, the average westward drifting velocity of the ROTI structure estimated from the ROTI maps in Figure 2 during 10:35-12:00 UT is ~150 m/s (~8 degree in longitude within ~1.5 hours), which is much smaller than the TID propagating velocity (~340 m/s) reported by the recent studies (Lin et al., 2022; Zhang et al., 2022). Therefore, the ROTI structures below ~33°N were more likely due to the extension of EPBs at the equator rather than locally triggered by TIDs through the Perkins instability.

With regard to the zonal drift of EPBs, it was westward in the present case that is opposite to the normal behavior of EPBs. The drift of EPBs was suggested to be driven by the thermospheric zonal winds which are usually eastward under geomagnetic quiet conditions (e.g., Muella et al., 2017). However, previous studies also reported westward-drifting EPBs during geomagnetic storms, when the zonal drift could be influenced by Hall drifts or the disturbance westward thermospheric winds (e.g., Abdu, 2012). If the westward drifts increase with latitude, the west-tilted EPB structures could be formed (e.g., Li et al., 2018), just as bubbles “A” and “B” in the present study.

# 5. Summary

We reported an unseasonal super EPB event at post sunset on 15 January 2022. The EPBs extended to middle latitudes up to ~33°N and caused intense L-band amplitude scintillations. The satellite signal fading depths caused by the EPBs were up to ~16 dB over a wide region from middle to low latitudes. Before the generation of EPBs, significant ionospheric perturbation and F layer height rise were observed around sunset, coinciding well with the arrival of waves produced by the Tonga volcano eruption. The ionospheric perturbation might act as a seeding source for the development of R-T instability. The eastward polarization electric field within the perturbation structure could further elevate the F layer to higher altitudes and enhance the growth rate of R-T instability. It was suggested that the ionospheric perturbations caused by the waves related with Tonga volcano eruption triggered the generation of super EPBs. Our results implicate that volcano eruption could produce large-scale ionospheric irregularities in the region more than ten thousand kilometers away and cause significant impact on ionospheric weather.

**Acknowledgments.**

This work was supported by the Project of Stable Support for Youth Team in Basic Research Field, CAS (YSBR-018), the National Natural Science Foundation of China (42020104002, 41727803, 42074190, 41904141), the “National key research and development program” (2017YFE0131400), the Solar-Terrestrial Environment Research Network (STERN) of Chinese Academy of Sciences, the International Partnership Program Of Chinese Academy of Sciences (Grant No.183311KYSB20200003), the Chinese Meridian

Project and the Informatization Plan of Chinese Academy of Sciences (Grant No. CAS-WX2021SF-0303, CAS-WX2021PY-0101).

The GNSS TEC and ionosonde data were obtained from the CMONOC, the Geophysics Center, National Earth System Science Data Center at BNOSE, IGGCAS (http://wdc.geophys.ac.cn/dbList.asp) and the Chinese Meridian Project (http://data.meridianproject.ac.cn). The IMF $B_z$, IEF $E_y$ and Dst index were obtained from the National Geophysical Data Center (https://www.ngdc.noaa.gov/ngdcinfo/onlineaccess.html). The magnetometer data were obtained from BCMT (http://www.bcmt.fr/data_download.php). The Swarm data is available from the European Space Agency (https://swarm-diss.eo.esa.int/). The AIRS data were obtained from the NASA Goddard Earth Sciences Data Information and Services Center (https://datapub.fz-juelich.de/slcs/airs/gravity_waves/). All the data used in this study are archived at the WDC for Geophysics, Beijing (https://doi.org/10.12197/2022GA003).

# References

Aa, E., Huang, W., Liu, S., Shi, L., Gong, J., Chen, Y., & Shen, H. (2015). A regional ionospheric TEC mapping technique over China and adjacent areas: GNSS data processing and DINEOF analysis. Science China Information Sciences, 58(10), 1–11. https://doi.org/10.1007/s11432-015-5399-2.

Abadi, P., Saito, S., Srigutomo, W. (2014). Low-latitude scintillation occurrences around the equatorial anomaly crest over Indonesia. Ann Geophys 32:7–17. https://doi.org/10.5194/angeo-32-7-2014

Abdu, M.A. (2012). Equatorial spread F/plasma bubble irregularities under storm time disturbance electric fields. J Atmos Solar Terr Phys 75–76:44–56. https://doi.org/10.1016/j.jastp .2011.04.024

Abdu, M.A. (2019). Day-to-day and short-term variabilities in the equatorial plasma bubble/spread F irregularity seeding and development. Prog Earth Planet Sci 6:11. https://doi.org/10.1186/s4064 5-019-0258-1

Alfonsi, L., Spogli, L., Pezzopane, M. et al. (2013). Comparative analysis of spread-F signature and GPS scintillation occurrences at Tucumán, Argentina. J Geophys Res Space Phys, 118:4483–4502. https://doi.org/10.1002/jgra.50378

Anderson, D., Anghel, A., Yumoto, K., Ishitsuka, M., & Kudeki, E. (2002). Estimating daytime vertical ExB drift velocities in the equatorial F‐region using ground‐based magnetometer observations. Geophysical Research Letters, 29(12), 1596. https://doi.org/10.1029/2001GL014562

Balan N., Maruyama T., Patra A. K. et al. (2018). A minimum in the latitude variation of spread-F at March equinox. Prog Earth Planet Sci 5:27. https://doi.org/10.1186/s4064 5-018-0180-y

Basu S., Mackenzie E., Basu S. (1988). Ionospheric constraints on VHF/UHF communication links during solar maximum and minimum periods. Radio Sci 23:363–372

Beutler, G., M. Rothacher, S. Schaer, T. A. Springer, J. Kouba, and R. E. Neilan (1999). The International GPS Service (IGS): An interdisciplinary service in support of Earth sciences, Adv. Space Res., 23, 631–635, doi:10.1016/S0273-1177(99)00160-X.

Buhari, S., Abdullah, M., Yokoyama, T. et al. (2017). Climatology of successive equatorial plasma bubbles observed by GPS ROTI over Malaysia. J Geophys Res Space Phys 122:2174–2184. https://doi.org/10.1002/2016J A0232 02

Cheng, K. and Huang, Y.-N. (1992). Ionospheric disturbances observed during the period of Mount Pinatubo eruptions in June 1991, Journal of Geophysical Research, 97, A11, 16995–17004. doi:10.1029/92JA01462.

D'Angelo, G., Piersanti, M., Diego, P., et al. (2021). Analysis of the August 14, 2018 plasma bubble by CSES-01 satellite. Il Nuovo Cimento C, 04(5), doi:10.1393/ncc/i2021-21118-2

Demyanov, V. V., Yasyukevich, Yu. V., Ishin, A. B., Astafyeva, E. I. (2012). Ionospheric super-bubble effects on the GPS positioning relative to the orientation of signal path and geomagnetic field direction. GPS Solut., 16(2), 181-189, doi: 10.1007/s10291-011-0217-9

Harding, B. J., Wu, Y. J., Yamazaki, P., Colin C. et al. (2022). Impacts of the January 2022 Tonga Volcanic Eruption on the Ionospheric Dynamo: ICON-MIGHTI and Swarm Observations of Extreme Neutral Winds and Currents, Geophysical Research Letters, 49, e2022GL098577. https://doi.org/10.1029/2022GL098577.

Hoffmann, L., X. Xue, and M. J. Alexander (2013). A global view of stratospheric gravity wave hotspots located with Atmospheric Infrared Sounder observations, J. Geophys. Res. Atmos., 118, 416-434, doi:10.1029/2012JD018658.

Fejer, B.G., Souza, J., Santos, A.S. et al. (2005). Climatology of F region zonal plasma drifts over Jicamarca. J Geophys Res, 110: A12310. https://doi.org/10.1029/2005J A011324

Friis-Christensen, E., Lühr, H., Knudsen, D., & Haagmans, R. (2008). Swarm-An Earth Observation Mission investigating Geospace. Advances in Space Research, 41 (1), 210-216. doi: 10.1016/j.asr.2006.10.008

Katamzi, Z. T., J. B. Habarulema, and M. Hernández-Pajares (2017), Midlatitude postsunset plasma bubbles observed over Europe during intense storms in April 2000 and 2001, Space Weather, 15, 1177–1190, doi:10.1002/2017SW001674.

Kelley M.C. (2009). The Earth's ionosphere: plasma physics and electrodynamics. International geophysics series, vol 43. Academic Press, San Diego

Keskinen, M. J., S. L. Ossakow, and B. G. Fejer (2003). Three-dimensional nonlinear evolution of equatorial ionospheric spread-F bubbles, Geophys. Res. Lett., 30(16), 1855, doi:10.1029/2003GL017418.

Li, G., Ning, B., Wang, C. et al. (2018). Storm-enhanced development of postsunset equatorial plasma bubbles around the meridian 120°E/60°W on 7–8 September 2017. J Geophys Res Space Phys, 123:7985–7998. https://doi.org/10.1029/2018J A025871

Li, G., Ning, B., Zhao, X., Sun, W., Hu, L., Xie, H., et al. (2019). Low latitude ionospheric TEC oscillations associated with periodic changes in IMF Bz polarity. Geophysical Research Letters, 46, 9379–9387. https://doi.org/10.1029/2019GL084428

Li, G., Ning, B., Otsuka, Y., Abdu, M., Abadi, P., Liu, Z., Spogli, L., Wan, W. (2021). Challenges to Equatorial Plasma Bubble and Ionospheric Scintillation Short-Term Forecasting and Future Aspects in East and Southeast Asia. Surveys in Geophysics. doi:10.1007/s10712-020-09613-5

Lin, J.-T., Rajesh, P. K., Lin, C. C. H., Chou, M.-Y., Liu, J.-Y., Yue, J. et al. (2022). Rapid Conjugate Appearance of the Giant Ionospheric Lamb Wave in the Northern Hemisphere After Hunga-Tonga

Volcano Eruptions. Earth and Space Science Open Archive, 18. doi: 10.1002/essoar.10510440.2

Ma, G. and Maruyama, T. (2006). A super bubble detected by dense GPS network at east Asian longitudes. Geophys Res Lett, 33:L21103. https://doi.org/10.1029/2006G L0275 12

Muella, M., Duarte-Silva, M.H., Moraes, A.O. et al. (2017). Climatology and modeling of ionospheric scintillations and irregularity zonal drifts at the equatorial anomaly crest region. Ann Geophys, 35:1201–1218. https://doi.org/10.5194/angeo -35-1201-2017

Otsuka, Y., Shiokawa, K., Ogawa, T., Wilkinson, P. (2002). Geomagnetic conjugate observations of equatorial airglow depletions. Geophys Res Lett., https://doi.org/10.1029/2002gl015347

Pi, X., Mannucci, A. J., Lindqwister, U. J., & Ho, C. M. (1997). Monitoring of global ionospheric irregularities using the Worldwide GPS Network. Geophysical Research Letters, 24(18), 2283–2286. doi:10.1029/97gl02273.

Piersanti M., Pezzopane M., Zhima Z. et al., (2020). Can an impulsive variation of the solar wind plasma pressure trigger aplasma bubble? A case study based on CSES, Swarm and THEMIS data, Advances in Space Research, https://doi.org/10.1016/j.asr.2020.07.046.

Seo, J., Walter, T., Chiou, T-Y., Enge, P. (2009). Characteristics of deep GPS signal fading due to ionospheric scintillation for aviation receiver design. Radio Sci, 44:RS0A16. https://doi.org/10.1029/2008rs004077

Schmidt, A., et al. (2014). Assessing hazards to aviation from sulfur dioxide emitted by explosive Icelandic eruptions, J. Geophys. Res. Atmos., 119, 14,180–14,196, doi:10.1002/ 2014JD022070.

Shi, J., Wang G., Reinisch B. et al. (2011). Relationship between strong range spread F and ionospheric scintillations observed in Hainan from 2003 to 2007. J Geophys Res 116:A08306. https://doi.org/10.1029/2011J A0168 06

Sun, W., Wu, B., Wu, Z., Hu, L., Zhao, X., & Zheng, J., et al. (2020). IONISE: An ionospheric observational network for irregularity and scintillation in East and Southeast Asia. Journal of Geophysical Research: Space Physics, 125, e2020JA028055. https://doi.org/10.1029/2020JA028055

Takahashi, H. et al. (2018). Equatorial plasma bubble seeding by MSTIDs in the ionosphere. Prog EarthPlanet Sci., 5:32. https://doi.org/10.1186/s40645-018-0189-2

Wang, C. (2010). New Chains of Space Weather Monitoring Stations in China, Space Weather, 8, S08001, doi:10.1029/2010SW000603.

Xiong, B., W. Wan, Y. Yu, and L. Hu (2016). Investigation of ionospheric TEC over China based on GNSS data, Adv. Space Res., 58(6), 867–877, doi:10.1016/j.asr.2016.05.033.

Xiong, C., Lühr, H., Sun, L., Luo, W., Park, J., & Hong, Y. (2019). Long-lasting latitudinal four-peak structure in the nighttime ionosphere observed by the Swarm constellation. Journal of Geophysical Research: Space Physics, 124, 9335–9347. https://doi. org/10.1029/2019JA027096

Zhang, S.-R., Vierinen, J., Aa, E., Goncharenko, L. P., Erickson, P. J., Rideout, W. et al. (2022). 2022 Tonga volcanic eruption induced global propagation of ionospheric disturbances via Lamb waves. Earth and Space Science Open Archive, 15. https://doi/abs/10.1002/ essoar.10510445.1

Zhao, X., Xie, H., Hu, L., Sun, W., et al. (2021). Climatology of equatorial and low-latitude F region kilometer-scale irregularities over the meridian circle around 120°E/60°W. GPS Solutions, 25(1). doi:10.1007/s10291-020-01054-2